\documentclass[twocolumn,aps]{revtex4}
\usepackage{amsmath}
\usepackage{amssymb}
\usepackage{graphicx}
\def\bnabla{\mbox{\boldmath $\nabla$}} 
\usepackage{dcolumn}
\usepackage{bm}
\usepackage{xcolor}

\begin{document}

\title{Superballistic boundary conductance and hydrodynamic transport in microstructures}

\author{O. E. Raichev}

\affiliation{Institute of Semiconductor Physics, NAS of Ukraine, Prospekt Nauki 41, 03028 Kyiv, Ukraine}

\date{\today}

\begin{abstract}

It is shown that the ideal boundary between a perfectly conducting electrode and electron 
liquid state acts as a contact whose conductance per unit area is higher than the fundamental 
Sharvin conductance by a numerical coefficient $2 \alpha$, where $\alpha$ is slightly 
smaller than unity and depends on the dimensionality of the system. If the boundary 
has a finite curvature, an additional correction to the boundary conductance appears, 
which is parametrically small as a product of the curvature by the electron-electron 
mean free path length. The relation of the normal current density to the voltage 
between the electrode and electron liquid represents itself a hydrodynamic boundary 
condition for current-penetrable boundary. Calculations of the conductance 
and potential distribution in microstructures by means of numerical solution of the 
Boltzmann equation show that the concept of boundary conductance works very good when 
the hydrodynamic transport regime is reached. The superballistic transport, when the 
device conductance is higher than the Sharvin conductance, can be realized in Corbino 
disk devices not only in the hydrodynamic regime, although requires that the electron-electron 
scattering must be stronger than the momentum-relaxing scattering. The theoretical results 
for Corbino disks are consistent with recent experimental findings.

\end{abstract}

\maketitle

\section{Introduction}

In the last years, there is a large progress in the studies of the hydrodynamic transport 
regime for electron gas in solid state conductors, when electron motion resembles the dynamics 
of viscous fluids \cite{gurzhi}-\cite{kumar}. This regime takes place under condition 
that the mean free path length with respect to momentum-conserving electron-electron (ee) 
scattering, $l_{e}$, is much smaller than the other characteristic lengths of the system, namely 
the transport mean free path length $l_{tr}$ describing momentum-relaxing (electron-impurity 
and electron-phonon) scattering and the lengths relevant to the conductor geometry. 
Since $l_e$ rapidly decreases with temperature, $l_e \propto T^{-2}$, the hydrodynamic 
regime can be reached by increasing electron temperature in high-mobility two-dimensional (2D) 
electron gas, where $l_{tr}$ is minimized. Among numerous fascinating manifestations of the 
hydrodynamic behavior, it has been found \cite{Hguo},\cite{RKkumar},\cite{ginzb},\cite{stern},\cite{kumar} 
that the conductance of microcontacts in the hydrodynamic regime can be higher 
than the conductance in the ballistic transport regime. This property, often called as 
superballistic transport, has been experimentally verified in narrow 2D constrictions (point 
contacts) with widths of the order of one micron, based on graphene \cite{RKkumar} and GaAs 
heterostructures \cite{ginzb}. Recently, the signatures of superballistic behavior have been 
also detected in high-quality graphene Corbino disks of several micron size \cite{kumar}. 

The fundamental upper bound of the ballistic conductance is given by the Sharvin conductance 
equal to the conductance quantum $e^2/2 \pi \hbar$ multiplied by the number of quantum channels 
able to carry the electrons through the contact (open channels). In the case of classical transport, 
this number is large and proportional to the area of the contact. The ideal ballistic Sharvin 
contact can be viewed as a hole in a thin non-penetrable wall separating two regions where 
electron gas stays in quasi-equilibrium characterized by different electrochemical potentials 
$U_1$ and $U_0$ \cite{sharvin}. A similar setup is realized in a point contact representing 
a smooth constriction between two regions, so the hole is identified with the 
narrowest place of this constriction. The Sharvin contact can be also created by placing 
a thin conducting layer, where electrons move ballistically, between two perfectly conducting 
electrodes, also called below as leads, as shown in Fig. 1 (a). It is assumed that 
the boundaries between the leads and the layer are ideal so that electrons pass them without 
backscattering. All such systems are characterized by the Sharvin conductance $G_S=I/U$, where 
$I$ is the total current passing through the contact and $U=U_1-U_0$ is the applied voltage. 
In the case of sharp contact boundaries (Fig. 1), it is also convenient to introduce the normal 
current density $j_n$ and the Sharvin conductance per unit square of the contact area (for 2D case, 
per unit length), ${\cal G}_{S}=j_n/U$. This conductance depends on the 
electron energy spectrum, electron density $n$, and temperature $T$.     

\begin{figure}[ht!]
\includegraphics[width=9cm,clip=]{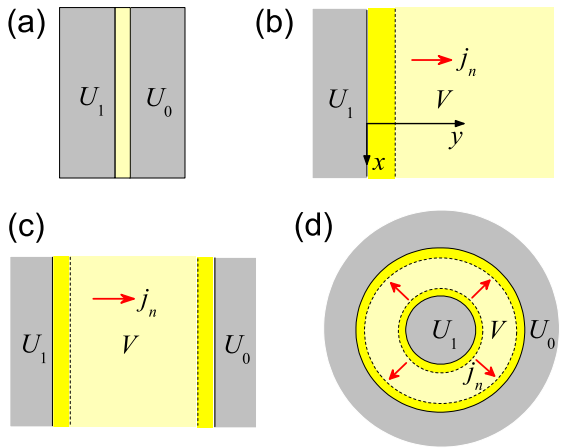}
\caption{\label{fig.1} (Color online) (a) A thin conducting layer, where electrons move 
ballistically, is sandwiched between two perfectly conducting electrodes with fixed 
electrochemical potentials, forming the Sharvin contact. (b) The contact between the 
electrode and the electron liquid staying at electrochemical potential $V$ spreads over 
the Knudsen layer (bright yellow), where the potential changes rapidly. The normal current 
flowing through the contact is indicated by the red arrow. (c,d) Devices of flat geometry 
and Corbino geometry with the contacts described above.}
\end{figure}

The theoretical explanation of the superballistic transport in constrictions 
\cite{Hguo},\cite{RKkumar},\cite{stern} is based on the fact that in the hydrodynamic regime 
the motion of electrons is collective and differs from the individual electron motion 
in the ballistic (Knudsen) regime. The Landauer interpretation of the contact conductance 
in the form given above is no longer valid for such hydrodynamic flow, though the concept 
of quantum channels still remains \cite{stern}. The number of these channels varies along 
the constriction, dropping down to the number of open channels in the narrowest place. 
In the process of motion, electron-electron scattering transfers carriers 
from the terminated channels, in which the ballistic electrons would scatter back, to the 
open channels, thereby helping them to pass the contact. To summarize, though the number of 
open channels remains the same, the electrons in the hydrodynamic regime use these channels 
more often than in the ballistic regime, so the conductance increases. The theory \cite{Hguo},\cite{stern} 
predicts that the conductance increases over the Sharvin one by a factor proportional to the 
constriction length divided by $l_e$. Thus, it was concluded \cite{stern} that, in terms 
of parameters, the conductance can be made arbitrary large.
 
The superballistic transport in the systems with sharp contact boundaries (Fig. 1) requires 
a different explanation. The main goal of this paper is to describe the transport mechanism in 
such systems and to find a specific conductance associated with the presence of the boundary 
and below referred to as the boundary conductance. The results are also applied to transport 
in microstructures. The consideration is based on the concept that the interface between 
the electrode and the electron liquid can be viewed as a quasiballistic Sharvin-like contact. 
Indeed, the hydrodynamic state away from the boundary is characterized by its own electrochemical 
potential $V$ and current density ${\bf j}$ governed by the continuity equation and the Navier-Stokes 
equation. The spatial variation of these quantities occurs on a much larger scale than the width of 
the Knudsen layer separating the hydrodynamic state from the lead, Fig. 1 (b). 
The linear relation between the normal inward current density and the potential drop $\Delta U$ between 
the lead and the electron liquid is written as 
\begin{eqnarray}
j_n= {\cal G} \Delta U.
\end{eqnarray}
The conductance per unit area of the boundary, ${\cal G}$, is considered in Sec. II. It 
is shown that the upper bound of ${\cal G}$ is equal to $2{\cal G}_S$, while the numerical calculations 
based on the Boltzmann equation formalism show that the ratio ${\cal G}/{\cal G}_S$ is slightly 
smaller than 2. The result depends on dimensionality as well as on the model of electron-electron 
collision integral. Thus, the contact between the lead and the electron system in the 
hydrodynamic state is superballistic. However, in contrast to the constriction described above, 
such a contact cannot provide arbitrary large conductance. Section III contains calculations of 
the conductance and potential profile for the devices of highly symmetric geometries shown in 
Fig. 1 (c,d) for different transport regimes. In particular, it is shown that the Corbino disk 
devices can demonstrate the superballistic conductance caused by strong electron-electron 
interaction, and not only in the hydrodynamic regime. Concluding remarks are given in Sec. IV. 

\section{Boundary conductance}

Since the interface between the perfectly conducting electrode and the hydrodynamic electron state 
spreads over the Knudsen layer width, which is much smaller than any other hydrodynamic length 
parameter of the system, it is sufficient to consider a homogeneous flat contact boundary and 
a constant normal current density $j_n$, as shown in Fig. 2 (b). The current density is constant 
according to the continuity equation $\bnabla \cdot {\bf j}=0$. A tangential current density can 
also be present in the system, but it is not relevant for calculations of the boundary conductance 
because the tangential current does not affect the normal current and potential distribution on 
the scale of the Knudsen layer width estimated as $l_e$. This statement remains true even in 
the presence of a magnetic field parallel to the boundary if this field is sufficiently weak 
so the cyclotron radius is much larger than $l_e$.   

The Knudsen layer exists because the distribution of electron momenta at the boundary is 
affected by injection of the electrons from the lead and, therefore, is different from the 
one in the hydrodynamic state. Thus, some space is needed to accommodate this distribution to 
the hydrodynamic form, which is achieved owing to electron-electron interaction. Description 
of the transport within the Knudsen layer requires solution of the Boltzmann kinetic equation
for the distribution function of electrons in the system, $f_{{\bf p}}({\bf r})$, where 
${\bf p}$ is the electron momentum and ${\bf r}$ is the coordinate. The energy spectrum of 
electrons, $\varepsilon=\varepsilon_{\bf p}$, is assumed to be isotropic. In the linear 
transport regime, it is convenient to present the distribution function as 
\begin{eqnarray}
f_{\bf p}({\bf r})= f_{\varepsilon}+ \delta f_{\bf p}({\bf r}) = f_{\varepsilon} 
-[g_{\bf p}({\bf r})- e\Phi({\bf r})] \partial_{\varepsilon} f,
\end{eqnarray}
where $f_{\varepsilon}$ is the equilibrium Fermi-Dirac distribution, 
$\partial_{\varepsilon} f \equiv \partial f_{\varepsilon}/\partial \varepsilon$ and 
$\Phi({\bf r})$ is the electrostatic potential that is equal to zero in equilibrium. 
The function $g_{\bf p}({\bf r})$ describes the non-equilibrium response. In particular, 
the current density is
\begin{eqnarray}
{\bf j}({\bf r})= e \int d \varepsilon D_{\varepsilon} \overline{{\bf v}_{{\bf p}} 
g_{\bf p}({\bf r})} (- \partial_{\varepsilon} f),
\end{eqnarray}   
where $D_{\varepsilon}$ is the density of states and ${\bf v}_{{\bf p}}$ is the 
group velocity. The overline symbol denotes averaging over the angles of momentum. 
The non-equilibrium part of electrochemical potential, $V({\bf r})$, is determined 
by the isotropic part of $g_{\bf p}$, denoted below as ${\overline g}$: 
\begin{eqnarray}
eV({\bf r})= \left< \overline{g_{\bf p}({\bf r})} \right>=\left< \overline{g}({\bf r}) \right>.
\end{eqnarray}
Here and below, the average of an arbitrary function $F$ over energy is defined as
$$
\left< F \right> \equiv \int d \varepsilon 
D_{\varepsilon} v_{\varepsilon} p_{\varepsilon} (- \partial_{\varepsilon} f) F_{\varepsilon}/nd,
$$
in view of the identity $nd=\int d \varepsilon D_{\varepsilon} v_{\varepsilon} 
p_{\varepsilon} (- \partial_{\varepsilon} f)$, where $v$ and $p$ are the absolute values 
of the group velocity and momentum, and $d$ is the dimensionality of the system. 
Equation (4) is consistent with the definition $V({\bf r})= 
\delta \mu({\bf r})/e + \Phi({\bf r})$, where $\delta \mu({\bf r})$ is the 
non-equilibrium part of the chemical potential. In the hydrodynamic regime, when 
$\overline{\delta f_{\bf p}({\bf r})} = -\delta \mu({\bf r}) \partial_{\varepsilon} f$, 
${\overline g}({\bf r})$ is energy-independent and ${\overline g}({\bf r})=eV({\bf r})$. 
For degenerate electron gas, the average over energy fixes the 
energy variable at the Fermi energy, $\varepsilon=\varepsilon_F$, so $eV({\bf r})$ is 
equal to ${\overline g}({\bf r})$ taken at the Fermi energy.  

The function $g_{\bf p}({\bf r})$ can be expanded in series of angular harmonics as 
\begin{eqnarray}
g_{{\bf p}}=\overline{g}+ g_{\alpha} c_{\alpha} + Q_{\alpha \beta}(c_{\alpha}c_{\beta}-
\delta_{\alpha \beta}/d) + \ldots,
\end{eqnarray}
where $\alpha$ and $\beta$ are the Cartesian coordinate indices (the repeated indices, 
by convention, imply summation over them), and ${\bf c}={\bf p}/p$ is the unit vector 
along the momentum. The vector $g_{\alpha} = d \overline{c_{\alpha} g_{\bf p}}$ and 
the tensor $Q_{\alpha \beta}$ depend on energy and coordinate. They are related to 
the drift velocity $u_{\alpha} = j_{\alpha}/en$ and to the momentum flux density 
tensor $\Pi_{\alpha \beta}$ as $u_{\alpha} = \langle g_{\alpha}/p \rangle$ 
and $\Pi_{\alpha \beta}=2 n \langle Q_{\alpha \beta} \rangle/(d+2)$. 
In the hydrodynamic regime, $g_{\alpha}=p u_{\alpha}$ and $Q_{\alpha \beta}=-p v \tau 
\frac{1}{2}( \nabla_{\beta} u_{\alpha} + \nabla_{\alpha} u_{\beta} )$, where $\tau$ 
is the relaxation time of the second angular harmonic of the distribution function, 
while the higher-order terms, denoted in Eq. (5) by the dots, should be neglected. 
The quantity $-\Pi_{\alpha \beta}$ describes the viscous stress tensor, 
$-\Pi_{\alpha \beta}=\eta ( \nabla_{\beta} u_{\alpha} + \nabla_{\alpha} u_{\beta} )$, 
where $\eta=n \langle p v \tau \rangle/(d+2)$ is the dynamic viscosity. 

Below, the axis normal to the boundary is chosen as $Oy$ and the boundary is placed 
at $y=0$, Fig. 1 (b). The electrons moving to the right and to the left are described 
by the distribution functions $f^{+}_{{\bf p}} = f_{{\bf p}}|_{p_y>0}$ and 
$f^{-}_{{\bf p}} = f_{{\bf p}}|_{p_y<0}$, respectively, defined in the momentum 
half-space $p_y=p \sin \varphi > 0$. The functions 
$g^{\pm}_{\bf p}$ are introduced in a similar way. The boundary condition at 
the left side, $y=0$, is written as
\begin{eqnarray}
\left. g^{+}_{\bf p}\right|_{y=0}=eU_1,
\end{eqnarray} 
which corresponds to representation of $f^{+}_{{\bf p}}$ as an isotropic Fermi-Dirac 
distribution characterized by the quasi-Fermi level of the left electrode. This is a particular 
case of the in-flow boundary condition applied to current-penetrable boundaries in kinetic theory 
\cite{abdallah,guo}, and its form is justified by the basic assumptions that the electron 
density and conductivity in the electrode are much larger than those in the electron system at 
$y>0$, and that the electrons pass the boundary without backscattering. The boundary condition 
at the right side, $y \gg l_e$, is derived from the known form of the distribution function, 
Eq. (5), taken in the hydrodynamic transport regime. Since the drift velocity is constant, 
the viscous stress at $y \gg l_e$ is zero, so that $g_{\bf p} = e V + p_{y} u_{y}$ and the 
boundary condition is
\begin{eqnarray}
\left. g^{-}_{\bf p}\right|_{y \gg l_e}=e V - (p/en) j_n \sin \varphi,
\end{eqnarray} 
since $u_{y}=j_n/en$ and $p_y=\pm p \sin \varphi$ for $\varphi \in [0,\pi]$ in $g_{\bf p}^{\pm}$. 
The normal current density $j_n$ is constant everywhere. According to Eq. (3), $j_n$ at $y=0$ is 
written as 
\begin{eqnarray}
j_n= e \int d \varepsilon D_{\varepsilon} v_{\varepsilon}
(- \partial_{\varepsilon} f) \overline{\left[ [ g^+_{\bf p}(0)-g^-_{\bf p}(0) ] 
\sin \varphi \right]}_+,
\end{eqnarray}   
where $g_{\bf p}$ is expressed in terms of $g^{\pm}_{\bf p}$ and $\overline{[\ldots]}_+$ 
denotes angular averaging limited by the half-space or half-plane where $p_y>0$. For three-dimensional (3D) 
systems, $\overline{[\ldots]}_+ \equiv \int_+ \frac{d \Omega}{4 \pi} \ldots = (4 \pi)^{-1} 
\int_0^{\pi/2} d \varphi \cos \varphi \int_0^{2 \pi} d \chi \ldots$, where $d \Omega$ is the 
differential of the solid angle and $\chi$ is the angle of the tangential component of momentum 
in the boundary plane $(xz)$. For 2D systems, $\overline{[\ldots]}_+ \equiv (2 \pi)^{-1} \int_0^{\pi} 
d \varphi \ldots$. Assuming that the form of $g^{-}_{\bf p}$ given by Eq. (7) remains valid all 
the way to the boundary $y=0$, which corresponds to the approximation that the left-moving 
electrons are not scattered in the Knudsen layer, one obtains $g^+_{\bf p}(0)-g^-_{\bf p}(0)=
e(U_1-V)+ (p/en)j_n \sin \varphi$. By applying Eq. (8) and noticing that the Sharvin conductance 
per unit area is 
\begin{eqnarray}
{\cal G}_S= e^2 \int d \varepsilon D_{\varepsilon} v_{\varepsilon}
(- \partial_{\varepsilon} f) \overline{[\sin \varphi]}_+ , 
\end{eqnarray}   
one can find that the potential term in the expression for $g^+_{\bf p}(0)-g^-_{\bf p}(0)$ produces 
the Sharvin current density ${\cal G}_S \Delta U$, where $\Delta U=U_1-V$, while the term 
proportional to $j_n$ is equal to $j_n/2$. Thus, one obtains Eq. (1) with
\begin{eqnarray}
{\cal G}/{\cal G}_S= 2,
\end{eqnarray}   
so the boundary conductance ${\cal G}$ is exactly twice larger than the Sharvin 
conductance. The derivation of Eq. (10) shows that the superballistic effect occurs because 
the boundary condition for the left-moving electrons includes the term proportional to the 
current density $j_n$. Without this term, one would obtain the usual Sharvin conductance
${\cal G}={\cal G}_S$. Physically, the presence of the current modifies the distribution 
function in the half-space $y>0$, in contrast to the perfectly conducting region at 
$y<0$, where such a modification is negligibly small. This increases electron 
transmission, because the backward, proportional to $f^-_{\bf p}$, component of the 
current decreases, so the difference $g^+_{\bf p}-g^-_{\bf p}$ gains a positive 
contribution proportional to $j_n$.

The result of Eq. (10) is approximate, because it is obtained under the assumption that the 
left-moving electrons are not scattered in the Knudsen layer. Thus, Eq. (10) gives the upper 
bound of the conductance $\cal G$ in Eq. (1). To obtain a more precise result, one should solve 
the kinetic equation with boundary conditions Eqs. (6) and (7). The kinetic equation in the 
linear transport regime is transformed to an equation for $g_{\bf p}$,
\begin{eqnarray}
{\bf v}_{{\bf p}} \cdot \bnabla g_{{\bf p}}({\bf r}) = J_{{\bf p}}({\bf r}),   
\end{eqnarray}
where the linearized collision integral is represented as $(- \partial_{\varepsilon} f) 
J_{{\bf p}}({\bf r})$. The electrostatic potential $\Phi$ does not appear explicitly in 
Eq. (11) because it is already included to the isotropic part of $g_{\bf p}({\bf r})$ according 
to Eq. (2). The collision-integral term $J_{{\bf p}}$ is written as a sum of momentum-relaxing (MR) 
and momentum-conserving electron-electron parts, $J_{\bf p}=J^{MR}_{\bf p} + J^{ee}_{\bf p}$. 
To specify them, the elastic relaxation-time approximation is used:
\begin{eqnarray} 
J^{MR}_{\bf p} = - \frac{g_{\bf p}-\overline{g}}{\tau_{tr}},
\end{eqnarray} 
and
\begin{eqnarray} 
J^{ee}_{\bf p} = -\frac{g_{\bf p}-\overline{g}-g_{\alpha} c_{\alpha} }{\tau_e}, 
\end{eqnarray} 
where $\tau_{tr}$ is the transport time and $\tau_{e}$ is the electron-electron scattering 
time. The corresponding lengths are introduced as $l_{tr}=v \tau_{tr}$ and $l_{e}=v \tau_{e}$. 
The single-time approximation Eq. (13) for electron-electron collision integral
is often used in theoretical calculations \cite{govorov},\cite{dejong},\cite{scaffidi},\cite{lucas1},\cite{lucas2},\cite{holder},\cite{raichev1},\cite{gupta},\cite{kumar} since 
it satisfies the principal properties of particle and momentum conservation, 
$\overline{J^{ee}_{\bf p}}=0$ and $\overline{{\bf p} J^{ee}_{\bf p}}=0$, 
and provides the easiest way to solve the kinetic equation. 

\begin{figure}[ht!]
\includegraphics[width=8.5cm,clip=]{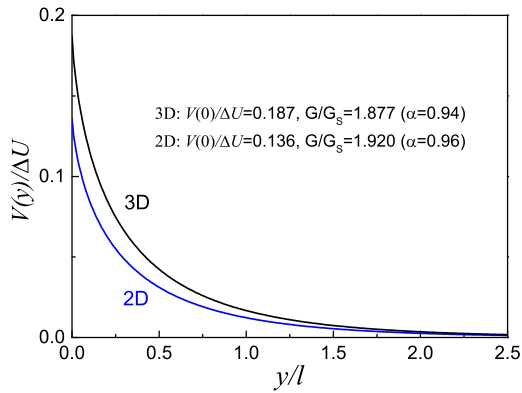}
\caption{\label{fig.2}(Color online) Distribution of electrochemical potential in the 
Knudsen layer for the systems of different dimensionalities. The drop of the potential 
across the Knudsen layer and the boundary conductance are indicated.}
\end{figure}

For the problem under the consideration [Fig. 1 (b)], Eq. (11) takes the form
\begin{eqnarray}
[\sin \varphi \nabla_y + l^{-1}]g_{\varphi}(y) = {\overline g}(y)/l + g_y \sin \varphi/l_e,
\end{eqnarray}
where $g_{\bf p}({\bf r})$ is written as $g_{\varphi}(y)$, $g_{y}=d \overline{g_{\varphi} 
\sin \varphi}$ is $y$-independent and proportional to the normal current density, and 
$l^{-1} \equiv l_{e}^{-1}+l_{tr}^{-1}$. Note that $l_e$, $l$, and $g_{\varphi}$ depend on 
electron energy $\varepsilon$. The solution of Eq. (14) satisfying the boundary conditions 
Eqs. (6) and (7) can be written through the integrals of ${\overline g}(y)$. Below, the 
zero point of $V(y)$ is chosen at $y \gg l$. Since the hydrodynamic transport in the bulk 
is considered, the momentum-relaxing scattering in the Knudsen layer is neglected, 
$l_{tr}^{-1} = 0$ and $l=l_e$. Owing to the elastic approximation for the collision 
integral, the angular averaging of Eq. (14) leads to the continuity equation $\nabla_y g_{y}=0$, 
which means that $g_y$ is a constant. Thus, $g_y$ is equal to its value at $y \gg l$ standing in 
the boundary condition Eq. (7), $g_y=p j_n/en$. With the use of identities ${\overline g}(y)=
\overline{[( g^+_{\varphi}+g^-_{\varphi} )]}_+$ and $g_y = d \overline{[( g^+_{\varphi} 
-g^-_{\varphi} ) \sin \varphi ]}_+$, the problem is finally reduced to an integral equation 
for ${\overline g}(y)$, 
\begin{eqnarray}
{\overline g}({\tilde y}) = e \Delta U S({\tilde y}) + \int_0^{\infty} d {\tilde y}' 
K({\tilde y},{\tilde y}') {\overline g}({\tilde y}'),
\end{eqnarray}
where ${\tilde y} = y/l$ and ${\tilde y}' = y'/l$ are the dimensionless coordinates,
\begin{eqnarray}
S({\tilde y}) = \zeta_0({\tilde y})- \zeta^2_1({\tilde y})/\zeta_2({\tilde y}), 
\end{eqnarray}
\begin{eqnarray}
K({\tilde y},{\tilde y}')= K_0({\tilde y},{\tilde y}') -  \frac{\zeta_1({\tilde y})}{\zeta_2({\tilde y})} 
K_1({\tilde y},{\tilde y}'),
\end{eqnarray}
\begin{eqnarray}
K_0({\tilde y},{\tilde y}')= \overline{[e^{-|{\tilde y}-{\tilde y}'|/\sin \varphi}/\sin \varphi]}_+, \nonumber \\
K_1({\tilde y},{\tilde y}')= {\rm sgn}({\tilde y}-{\tilde y}') \overline{[e^{-|{\tilde y}-{\tilde y}'|/
\sin \varphi} ]}_+, 
\end{eqnarray}
and 
\begin{eqnarray}
\zeta_k({\tilde y})= \overline{[\sin^k \varphi e^{-{\tilde y}/\sin \varphi}]}_+ .
\end{eqnarray}
Numerical solution of Eq. (15) allows one to find the function ${\overline g}(y)$ determining 
the potential distribution $V(y)$ and the current density $j_n$. The latter 
is given by Eq. (1), where, in contrast to Eq. (10), the conductance is smaller than $2 {\cal G}_S$:
\begin{eqnarray}
\frac{{\cal G}}{{\cal G}_S}=2 \alpha, ~ \alpha=1 - \frac{1}{\zeta_1(0)} \left< \left< \int_0^{\infty} d \tilde{y} 
\zeta_0({\tilde y}) \frac{{\overline g}({\tilde y})}{e \Delta U} \right>\right>.
\end{eqnarray}
In this equation, $\left< \left< \ldots \right> \right> \equiv \langle p^{-1} \ldots 
\rangle / \langle p^{-1} \rangle$ denotes the modified averaging over energy. 

Since Eq. (14) contains only one length parameter $l=l_e$, the dimensionless ratio 
${\overline g}({\tilde y})/e \Delta U$ is a numerical function of the dimensionless 
coordinate ${\tilde y}$. This property is directly seen from Eq. (15). 
The function ${\overline g}(y)$ depends on energy because of the energy 
dependence of the relaxation length $l$. However, the integral standing in 
Eq. (20) is an energy-independent numerical constant that is not affected 
by the averaging over energy. Therefore, the coefficient $\alpha$ describing the lowering of 
the ratio ${\cal G}/{\cal G}_S$ below its upper bound of 2 is independent of temperature, electron 
energy spectrum, and electron-electron scattering rate. The calculation gives $\alpha 
\simeq 0.96$ and $\alpha \simeq 0.94$ for 2D and 3D cases, respectively. The reduction of 
the boundary conductance caused by electron collisions in the Knudsen layer appears to be 
relatively small, and the conductance remains superballistic.  

For degenerate electron gas, when the electrochemical potential $V(y)$ is equal to 
${\overline g}(y)/e$ at the Fermi energy $\varepsilon_F$, and $l$ is fixed by its 
value at $\varepsilon_F$, the profile of $V(y)$ is shown in Fig. 2. The plots 
for 2D and 3D cases are similar and demonstrate that the main decrease of the 
potential occurs already at $y<l$. The 
total potential drop at the contact, $\Delta U$, from the hydrodynamic point of view, can be 
described in terms of the pressure jump under injection of liquid between two different reservoirs. 
It includes a sharp jump of the electrochemical potential at the boundary and a smooth decrease 
within the Knudsen layer. The sharp jump is a consequence of chemical potential drop caused by the
ballistic transfer of electrons between different media, whereas the smooth decrease appears 
because of the scattering of electrons within the Knudsen layer. The electrostatic potential $\Phi(y)$, 
which is related to the non-equilibrium density $\delta n$ according to Poisson's equation, 
does not show a sharp jump, although it follows the electrochemical potential profile within the 
spatial scale of the screening length. The magnitude of the smooth part of the potential drop is 
relatively small. The normalized magnitude, $V(0)/\Delta U$, similar as $\alpha$, depends only 
on the dimensionality.

\begin{figure}[ht!]
\includegraphics[width=8.5cm,clip=]{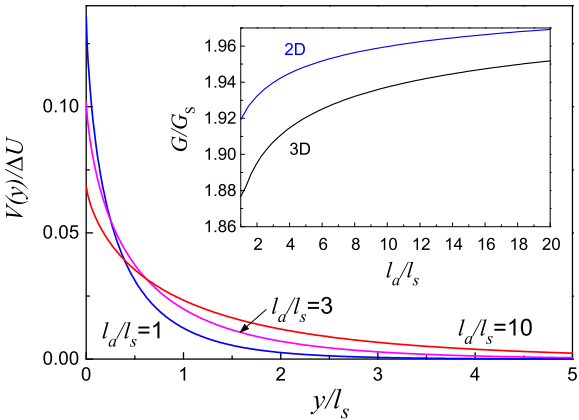}
\caption{\label{fig.3}(Color online) Distribution of electrochemical potential in the 
Knudsen layer for 2D systems in the case of modified relaxation-time approximation 
for the collision integral described by the two-time model of Eq. (21). The inset 
shows the dependence of the boundary conductance on the ratio of relaxation 
lengths for antisymmetric and symmetric parts of the distribution function.}
\end{figure}

The universality of the ratio ${\cal G}/{\cal G}_S$ described above is a consequence of 
the chosen model of the electron-electron collision integral, Eq. (13). A more careful 
consideration, even within the elastic approximation, suggests that this model is 
oversimplified because is describes the relaxation of all angular harmonics of 
electron distribution by a single time, $\tau_e$. In particular, it has been shown 
\cite{gurzhi1,gurzhi2} and later emphasized \cite{ledwith,ledwith1} that, due to the kinematic 
constraints, the main contribution to the electron-electron collision integral at low 
temperatures comes from the head-on collisions that cause relaxation of the momentum-symmetric 
part of the electron distribution while leaving the antisymmetric part intact. As a result, 
the symmetric, $f_{\bf p}^s=(f_{\bf p}+f_{-{\bf p}})/2$, and antisymmetric, $f_{\bf p}^a=
(f_{\bf p}-f_{-{\bf p}})/2$, parts of electron distribution are expected to relax with 
different times, $\tau_s$ and $\tau_a$, respectively, and $\tau_a$ should be considerably
larger than $\tau_s$ if $T$ is much smaller than the Fermi energy $\varepsilon_F$. 
The simplest way to account this difference is to introduce the elastic two-time 
model \cite{gurzhi1}, 
\begin{eqnarray} 
J^{ee}_{\bf p} = -\frac{g^s_{\bf p}-\overline{g}}{\tau_s}-\frac{g^a_{\bf p}-g_{\alpha} c_{\alpha} }{\tau_a}, 
\end{eqnarray} 
that generalizes Eq. (13). This model can be applied as well to the problem of boundary conductance.
Then, instead of Eq. (14), one has two equations  
\begin{eqnarray}
\sin \varphi \nabla_y g^a_{\varphi}(y) + g^s_{\varphi}(y)/l_s = {\overline g}(y)/l_s, \nonumber \\
\sin \varphi \nabla_y g^s_{\varphi}(y) + g^a_{\varphi}(y)/l_a = g_y \sin \varphi/l_a,
\end{eqnarray}
where $l_s=v \tau_s$, $l_a=v \tau_a$, and $l_{tr}^{-1}=0$ is already implied. It is sufficient to define 
the functions $g^s_{\varphi}(y)$ and $g^a_{\varphi}(y)$ in the region $p_y>0$, where $\sin \varphi$ is 
positive. Then, according to the symmetry of the problem, 
$g^{\pm}_{\varphi}(y)=g^s_{\varphi}(y) \pm g^a_{\varphi}(y)$. 
Solution of Eq. (22) with the boundary conditions Eqs. (6) and (7) leads to an integral equation of the 
same form as Eq. (15), where the dimensionless coordinates are now defined as ${\tilde y} = y/l_s$ and 
${\tilde y}' = y'/l_s$. The term $S$ in this equation is modified as
\begin{eqnarray}
S({\tilde y}) =\frac{2}{\beta+1} \left[\zeta_0({\tilde y})- \frac{\zeta^2_1({\tilde y})}{\zeta_2({\tilde y})}\right], 
\end{eqnarray}
with $\beta=\sqrt{l_a/l_s}=\sqrt{\tau_a/\tau_s}$ and 
\begin{eqnarray}
\zeta_k({\tilde y})= \overline{[\sin^k \varphi e^{-{\tilde y}/\beta \sin \varphi}]}_+. 
\end{eqnarray} 
The modified kernel $K$ is given by Eq. (17) with
\begin{eqnarray}
K_0({\tilde y},{\tilde y}')=\beta^{-1} \overline{[e^{-|{\tilde y}-{\tilde y}'|/\beta \sin \varphi}/\sin \varphi]}_+ \nonumber \\
+ \beta^{-1} \frac{\beta-1}{\beta+1} \overline{[e^{-({\tilde y}+{\tilde y}')/\beta \sin \varphi}/\sin 
\varphi]}_+ , \nonumber \\
K_1({\tilde y},{\tilde y}')=
\beta^{-1} {\rm sgn}({\tilde y}-{\tilde y}') \overline{[e^{-|{\tilde y}-{\tilde y}'|/\beta 
\sin \varphi} ]}_+ \nonumber  \\
+ \beta^{-1} \frac{\beta-1}{\beta+1} \overline{[e^{-({\tilde y}+{\tilde y}')/\beta \sin \varphi}]}_+,
\end{eqnarray}
and with $\zeta_k({\tilde y})$ of Eq. (24). Finally, Eq. (20) is modified by multiplying 
the integral term by $\beta^{-1}$ and using there $\zeta_0({\tilde y})$ of Eq. (24). If $l_a=l_s=l$ 
($\beta=1$), the problem is reduced to the one described by Eqs. (15)-(20). The presence of the 
energy-dependent factor $\beta$ in the modified equations makes the energy averaging in the modified 
Eq. (20) essential, in contrast to the initial Eq. (20). However, if energy dependence of $\beta$ is neglected, 
the boundary conductance ${\cal G}$ again does not depend on temperature, energy spectrum, and scattering 
time $\tau_s$, although the dependence of ${\cal G}$ on $\beta$ remains. The results of numerical calculations 
of the potential distribution for degenerate 2D electron gas is shown in Fig. 3, along with the dependence 
of the conductance on $l_a/l_s$. Increasing the ratio $l_a/l_s$ brings the conductance closer to its upper 
bound, which is expectable since the total relaxation rate $\tau_s^{-1}+\tau_a^{-1}$ decreases. However, 
this also increases the width of the Knudsen layer. The transport properties of the hydrodynamic state 
beyond the Knudsen layer are governed only by the length $l_s$ entering the expression for dynamic 
viscosity, $\eta=n \langle p l_s \rangle/(d+2)$. Indeed, the viscosity is determined by the relaxation 
time of the second angular harmonic of the distribution function, and this harmonic belongs to the 
symmetric set including all even harmonics. Thus, in contrast to the boundary conductance, the 
hydrodynamic transport in the bulk is not sensitive to whether the collision integral is given 
by Eq. (13) or Eq. (21), provided that both $l_s$ and $l_a$ are smaller than the other 
characteristic lengths. 

The above consideration shows that electron scattering in the Knudsen layer produces a numerically 
small relative deviation of the boundary conductance from the approximate result of Eq. (10). 
The use of the improved model of electron-electron collision integral makes this deviation even 
smaller. This observation brings much credit to the method applied in derivation of Eq. (10). A similar 
method, applied to calculation of the slip length in the hydrodynamic boundary condition for tangential 
current, also produces the results that are very close to those obtained in the detailed calculations 
taking into account electron collisions in the Knudsen layer \cite{raichevu}. Below, this method is 
applied in order to find the correction to the boundary conductance caused by a finite curvature 
of the boundary. The consideration is limited to 2D systems. The hydrodynamic description of the boundary 
implies that possible variation of the curvature occurs on a large scale compared to the Knudsen layer 
width. Thus, one can consider a piece of boundary with a constant curvature $C$ which is positive for 
convex boundary and negative for concave boundary. In contrast to the case of flat 
boundary, the normal current density $j_n$ near the curved boundary depends on the normal (radial) 
coordinate, in order to satisfy the current continuity, see also Sec. III. As a result, the viscous 
stress is generated in the bulk, so the distribution function there acquires an additional contribution 
proportional to $Q_{\alpha \beta}$, see Eq. (5). Calculating $Q_{\alpha \beta}$ in the hydrodynamic 
regime as described after Eq. (5), one obtains, instead of Eq. (7),
\begin{eqnarray}
g^{-}_{\bf p}=e V - (p/en) j_n \sin \varphi - C (p v \tau/en) j_n  \cos 2 \varphi.
\end{eqnarray} 
where $\varphi$ denotes the angle of momentum with respect to the tangent to the boundary, which 
is reduced to the angle $\varphi$ of the previous consideration if the curvature goes to zero. 
Applying Eq. (26) in Eq. (8), one gets 
\begin{eqnarray}
\frac{{\cal G}}{{\cal G}_S}= \frac{2}{1 + (4/3 \pi) C \langle v \tau \rangle},
\end{eqnarray}   
which generalizes Eq. (10) to the case of 2D systems with curved contact boundaries. The convex 
boundary decreases the conductance, while the concave boundary increases it. Within the 
simplest model of the collision integral given by Eqs. (12) and (13), the length $v \tau$ is 
identified with $l \simeq l_e$, whereas in the modified model of Eq. (21) 
$v \tau=l_s$. At low temperatures, when the electron gas is degenerate, this length
can be expressed through the viscosity according to $\langle v \tau \rangle= 4 \eta/n p_F$, 
where $p_F$ is the Fermi momentum. The correction due to the curvature is parametrically small, because 
the product $|C| l$ must be small in the hydrodynamic transport regime. Nevertheless, this 
correction may become more important than the small numerical correction proportional to the 
difference $1-\alpha$. To take into account both these corrections, one may use the following 
expression:
\begin{eqnarray}
\frac{{\cal G}}{{\cal G}_S}= \frac{2}{\alpha^{-1} + (4/3 \pi) C \langle v \tau \rangle}.
\end{eqnarray}
When applying Eq. (28), one should neglect the terms containing double smallness 
$\propto (1-\alpha) C \langle v \tau \rangle$, since such terms are beyond the accuracy 
of the approximation.

\section{Transport in miscrostructures}

The results of the previous section can be applied for calculation of the conductance 
of various microstructures with contacts. In this section, the simplest examples of 
two-terminal devices are considered, where the electron transport can be described as well
by solving the Boltzmann kinetic equation in a straightforward way. The kinetic equation 
approach makes it possible to study different transport regimes and transitions between them, 
and to link such results to those following from the above theory. The consideration below is 
limited to 2D systems, and the electron-electron collision integral is described by Eq. (13).
The case of degenerate electron gas is studied, so the characteristic lengths $l_e$ and 
$l_{tr}$ appearing in the theory correspond to $\varepsilon=\varepsilon_F$ and have 
the direct meaning of the mean free path lengths.  

Consider first the system of flat geometry shown in Fig. 1 (c). Although it is implied that the 
device has a finite width $W$ along $Ox$, the presence of the side walls is not essential as 
it is assumed that electrons are specularly reflected from these walls. In this case, the device 
behaves like an infinitely wide one, and its conductance in the ballistic limit is equal to the 
Sharvin conductance. The current flows along $Oy$ and the current density $j_n=j_y$ does not 
depend on coordinates, whereas the electrochemical potential depends only on $y$. In the hydrodynamic 
regime, specular reflection is equivalent to the no-stress boundary condition, $\nabla_x j_y=0$ 
at the side walls, which is satisfied automatically because the current density is constant. As 
there are no current density gradients, the linearized Navier-Stokes equation describing the 
current in the bulk of the system is reduced merely to the ohmic relation 
between the current density and the gradient of electrochemical potential in the bulk: 
$j_y=-\sigma_0 \nabla_y V(y)$, where $\sigma_0$ is the Drude conductivity. Thus, $\nabla_y V(y)$ 
is constant, and the total resistance of the system is determined as a resistance in series, formed 
by the sum of two equal boundary resistances, ${\cal R}_0={\cal R}_1=(2 \alpha G_S)^{-1}$, and 
the bulk ohmic resistance. The latter is limited by momentum-relaxing scattering and is equal 
to ${\cal R}_{bulk}=\sigma_0^{-1} L/W$, where $L$ is the distance between the contacts. In 2D 
systems with degenerate electron gas, $\sigma_0=
e^2 n l_{tr}/\hbar k_F$, $G_{S}={\rm g} e^2 k_F W/2 \pi^2 \hbar$, and $n={\rm g}k_F^2/4 \pi$, where 
$k_F$ is the Fermi wavenumber and ${\rm g}$ is the degeneracy factor of electron states (e.g., 
${\rm g}=2$ in GaAs and ${\rm g}=4$ in graphene). Then ${\cal R}_{bulk}=G_S^{-1} 2 L/\pi l_{tr}$, 
and the total resistance of the device is
\begin{eqnarray}
{\cal R}=G_S^{-1} \left[\frac{1}{\alpha} + \frac{2}{\pi} \frac{L}{l_{tr}} \right],  
\end{eqnarray} 
where $\alpha \simeq 0.96$ according to the results of the previous section. Since $\alpha<1$, 
such a device can never be superballistic, even if the momentum-relaxing scattering is absent. 
Thus, the momentum-conserving scattering alone increases the resistance of the system shown in 
Fig. 1 (c), making it larger than the Sharvin resistance $G_S^{-1}$. 

The expression for the resistance in the form similar to Eq. (29) remains valid for arbitrary 
$l_e$, $l_{tr}$, and $L$. This general case can be investigated by solving the kinetic equation. 
The non-equilibrium part of the distribution function is again governed by Eq. (14), with the 
boundary condition at $y=0$ given by Eq. (6). Accordingly, at the right side $y=L$, the boundary 
condition is $\left. g^{-}_{\bf p}\right|_{y=L}=eU_0$. If the point of zero electrochemical 
potential is chosen in the middle of the device, $y=L/2$, and the total applied voltage
is defined as $U=U_1-U_0$, the whole set of the boundary conditions is written as
\begin{eqnarray}
g^{+}_{\varphi}(0)=eU/2,~g^{-}_{\varphi}(L)=-eU/2.
\end{eqnarray}  
The solution of the Cauchy problem defined by Eq. (14) and the boundary conditions of Eq. (30) 
is facilitated by the observation that $g_{\varphi}(y)$ can be represented in the form 
$g_{\varphi}(y)=h_{\varphi}(y) - (y-L/2)g_y/l_{tr}$, where $h_{\varphi}(y)$ 
satisfies Eq. (14) with $l_e=l$, i.e., with zero momentum-relaxing scattering, $l_{tr} 
\rightarrow \infty$. Accordingly, the boundary conditions for $h_{\varphi}$ are modified 
by the formal substitution $eU \rightarrow eU-g_yL/l_{tr}$ so that $I=GU=G^*(U-g_yL/el_{tr})$, 
where $G^*$ is the effective conductance for the problem with zero momentum-relaxing scattering 
rate $1/l_{tr}$ and enhanced momentum-conserving scattering rate, $1/l_e \rightarrow 1/l_e+
1/l_{tr} \equiv 1/l$. Since $g_y=(2/\pi) e j_y/{\cal G}_S=(2/\pi) (G/G_S) eU$, the total 
resistance ${\cal R}=1/G$ is given by Eq. (29), where $\alpha$ is replaced by $G^*/G_S$. The 
problem of finding $G^*$ is described by Eq. (14) with $l_e=l$ and boundary conditions Eq. (30). 
It is reduced to solution of an integral equation similar to Eq. (15), see Appendix A. The 
ratio $G^*/G_S$ depends on a single parameter, the Knudsen number ${\rm K}$, defined here as 
${\rm K} \equiv l/L$. In the ballistic limit, ${\rm K} \rightarrow \infty$, $G^*=G_S$. 
In the hydrodynamic limit, ${\rm K} \rightarrow 0$, $G^*/G_S=\alpha \simeq 0.96$.

\begin{figure}[ht!]
\includegraphics[width=8.5cm,clip=]{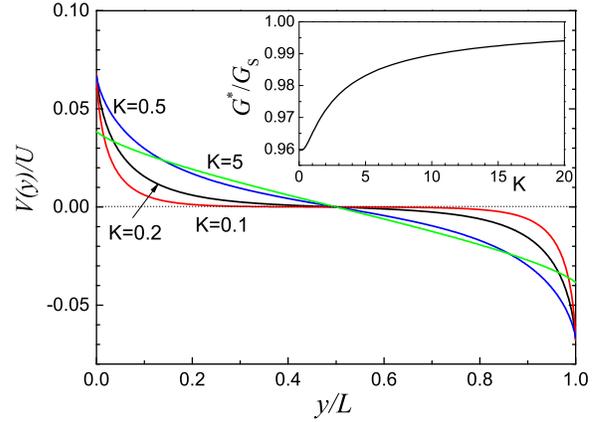}
\caption{\label{fig.4}(Color online) Distribution of electrochemical potential along the 
2D device with flat boundaries shown in Fig. 1 (c), in the absence of momentum-relaxing scattering. 
The inset shows the dependence of the conductance on the Knudsen number ${\rm K} \equiv l/L$.}
\end{figure}

In summary, the resistance of the system in Fig. 1 (c) in the general case is given by 
\begin{eqnarray}
{\cal R}= {\cal R}^* + \frac{2}{\pi} G_S^{-1} \frac{L}{l_{tr}},  
\end{eqnarray} 
where ${\cal R}^* \equiv 1/G^*$ is determined from Eqs. (A3)-(A7). In the hydrodynamic transport 
regime, when $l_e \ll l_{tr}$ and $l_e \ll L$, the effective resistance ${\cal R}^*$ is described 
as a sum of two boundary resistances and is equal to $(\alpha G_S)^{-1}$, so Eq. (31) is reduced to 
Eq. (29). The ratio $G^*/G_S$ changes from 0.96 to 1 as the length $l$ increases. The dependence 
of this ratio on the Knudsen number is shown in Fig. 4, which also demonstrates the potential 
profiles for different ${\rm K}$ in the absence of momentum-relaxing scattering. In the hydrodynamic 
regime, ${\rm K} \ll 1$, half of the total potential drops near each boundary, while in the bulk 
the potential is constant. When ${\rm K}$ increases above 1, the potential profile approaches 
a linear one, and the potential drop caused by the momentum-conserving scattering decreases.

Consider now the 2D Corbino disk system, Fig. 1 (d), where the conducting layer is placed between 
the circular contact boundaries with radii $R_1$ (inner) and $R_0$ (outer). In the absence of 
magnetic field, only the radial flow with current density $j$ is present, and all macroscopic 
quantities depend on the radial coordinate $r$. The continuity equation $\bnabla \cdot {\bf j}=0$ 
assumes the form
\begin{eqnarray}
\nabla_{r} j(r) + j(r)/r=0, 
\end{eqnarray}   
so that $j(r) = I/2 \pi r$, where $I$ is the total 
current. This property leads to disappearance of the viscous force, which means that the term 
proportional to the viscosity does not enter the Navier-Stokes equation \cite{shavit}, similar as 
in the case of the device with flat boundaries studied above. However, due to finite curvature of 
the boundaries, the current generates the viscous stress not only in the Knudsen layers, but also 
in the bulk of the system. The absence of the viscous term reduces the Navier-Stokes equation to the 
ohmic relation $j(r)=-\sigma_0 \nabla_{r} V(r)$ leading to the well known logarithmic dependence 
of the electrochemical potential in the bulk: $V(r)=V(R_0) - I \ln(r/R_0)/2 \pi \sigma_0$, so 
the bulk resistance is ${\cal R}_{bulk} = \ln(R_0/R_1)/2 \pi \sigma_0$. The resistances 
of the outer and inner boundaries are ${\cal R}_{0}=(2 \pi R_0 {\cal G})^{-1}$ and ${\cal R}_{1}
=(2 \pi R_1 {\cal G})^{-1}$, because the contact widths are equal to circumference 
of the boundaries. The curvatures of these boundaries are $-1/R_0$ and $1/R_1$. Thus, according 
to Eq. (28) with $\langle v \tau \rangle=l$, the sum of the resistances is 
\begin{eqnarray}
{\cal R}^*={\cal R}_{0}+{\cal R}_{1}=G_S^{-1} \left[\frac{b+1}{2 b \alpha} + \frac{2(b^2-1)}{3 \pi b} 
{\rm K} \right],
\end{eqnarray}
where $b \equiv R_0/R_1$. The Knudsen number is defined here as ${\rm K} \equiv l/R_0$. 
The ballistic conductance of the Corbino disk device is equal to the Sharvin 
conductance \cite{kumar}
\begin{eqnarray}
G_{S}= 2 \pi R_1 {\cal G}_S={\rm g} e^2 k_F R_1/\pi \hbar,
\end{eqnarray} 
which is proportional to the circumference of the inner contact as the ballistic electron flow 
is limited by the smallest circumference. One can also derive Eq. (34) from the kinetic equation 
in the ballistic limit (see Appendix B). By adding the resistances in series, one obtains the 
total resistance of the device in the hydrodynamic transport regime:
\begin{eqnarray}
{\cal R}={\cal R}^*  + \frac{2}{\pi} G_S^{-1} \frac{R_1 \ln b}{l_{tr}}.  
\end{eqnarray} 
In contrast to the device with flat boundaries, the Corbino 
disk device can be superballistic, i.e., ${\cal R}$ can be smaller than the Sharvin resistance $G_S^{-1}$.
The contribution to ${\cal R}^*$ due to boundary curvatures, given by the second term in the right-hand side 
of Eq. (33), coincides with the resistance calculated in Ref. \cite{shavit} up to a numerical coefficient 
$2/3$. Note, however, that the authors of Ref. \cite{shavit} considered a problem when this contribution
was the main one, while in the present theory it represents a small correction to the total resistance. 
If the momentum-relaxing scattering is absent, the total resistance is equal to the sum of the boundary 
contact resistances, ${\cal R}={\cal R}^*$, and the potential in the bulk of the device is constant, $V(r)=
V_{bulk}$ \cite{shavit}. Since $V_{bulk}={\cal R}_{0} I$,  
\begin{eqnarray}   
\frac{V_{bulk}}{U} =  \frac{1-4 {\rm K}/3 \pi}{(b+1)(1+4 {\rm K} (b-1)/3 \pi)}.
\end{eqnarray}
In the hydrodynamic limit, ${\rm K} \rightarrow 0$, Eqs. (33) and (36) assume the simple 
forms ${\cal R}^* = G_S^{-1}(b+1)/2 b \alpha$ and $V_{bulk}/U=1/(b+1)$. The inequality ${\rm K} \ll 1$ 
is the necessary condition for the validity of hydrodynamic description of the transport. 
The sufficient conditions are $l \ll {\rm min} \{R_1,R_0-R_1\}$ and $l_e \ll l_{tr}$.

Equation (35) with properly redefined ${\cal R}^*$ remains valid in the case of arbitrary $l_e$, $l_{tr}$, $R_0$, 
and $R_1$. To prove this statement, one needs to consider the kinetic equation. In the Corbino geometry, 
the function $g_{\bf p}({\bf r})$ can be written as $g_{\varphi}(r)$, where $\varphi$ now denotes the 
angle of momentum with respect to the tangent to the inner boundary. The angle $\varphi$ has the same 
meaning as before if the radial direction is identified with the $Oy$ axis. Equation (11) for 
$g_{\varphi}(r)$ assumes the form 
\begin{eqnarray}
\sin \varphi \nabla_{r} g_{\varphi}(r) + \frac{1}{r} \cos \varphi \frac{\partial 
g_{\varphi}(r)}{\partial \varphi} +\frac{g_{\varphi}(r)}{l} 
\nonumber \\
=\frac{\overline{g}(r)}{l} + \frac{A \sin \varphi}{l_e r},
\end{eqnarray}   
where the radial (the only nonzero) component of the vector ${\bf g}$ is written as $A/r$, 
since it is proportional to the current density. The constant $A$ is expressed as $A=I/\pi e D_F v_F$, 
where $D_F$ and $v_F$ are the density of states and group velocity at the Fermi level. Similar as above, 
it is convenient to decompose $g_{\varphi}(r)$ into $g^+_{\varphi}=g_{\varphi}$ and $g^-_{\varphi} = 
g_{2 \pi-\varphi}$ defined in the angular interval $\varphi \in [0,\pi]$ and describing the particles 
moving from the center and towards the center, respectively. Then, the in-flow boundary conditions 
for $g_{\varphi}(r)$ are
\begin{eqnarray}
g^{+}_{\varphi}(R_1)=eU,~g^{-}_{\varphi}(R_0)=0,
\end{eqnarray}  
where the point of zero potential is chosen at the outer electrode. The Boltzmann kinetic 
equation with boundary conditions of Eq. (38) has been used in the analysis of experimental 
data for the Corbino disk with $R_0/R_1=4.5$ in Ref. \cite{kumar}. The function $g_{\varphi}(r)$ 
is representable in the form $g_{\varphi}(\rho)=h_{\varphi}(r) - A \ln(r/R_0)/l_{tr}$, 
where $h_{\varphi}(r)$ satisfies Eq. (37) with $l_e=l$. Accordingly, the boundary conditions 
for $h_{\varphi}$ are modified by the substitution $eU \rightarrow eU - A \ln b /l_{tr}$. This 
leads to Eq. (35), where the effective resistance ${\cal R}^*$ is now found by solving the problem 
described by Eqs. (37) and (38) with $l_e=l$, which is formally equivalent to the transport 
problem with zero momentum-relaxing scattering rate and enhanced momentum-conserving scattering 
rate, $1/l_e \rightarrow 1/l_e+1/l_{tr} \equiv 1/l$ \cite{kumar}. This problem is solved by the 
method of characteristics as described in Appendix B.

\begin{figure}[ht!]
\includegraphics[width=8.5cm,clip=]{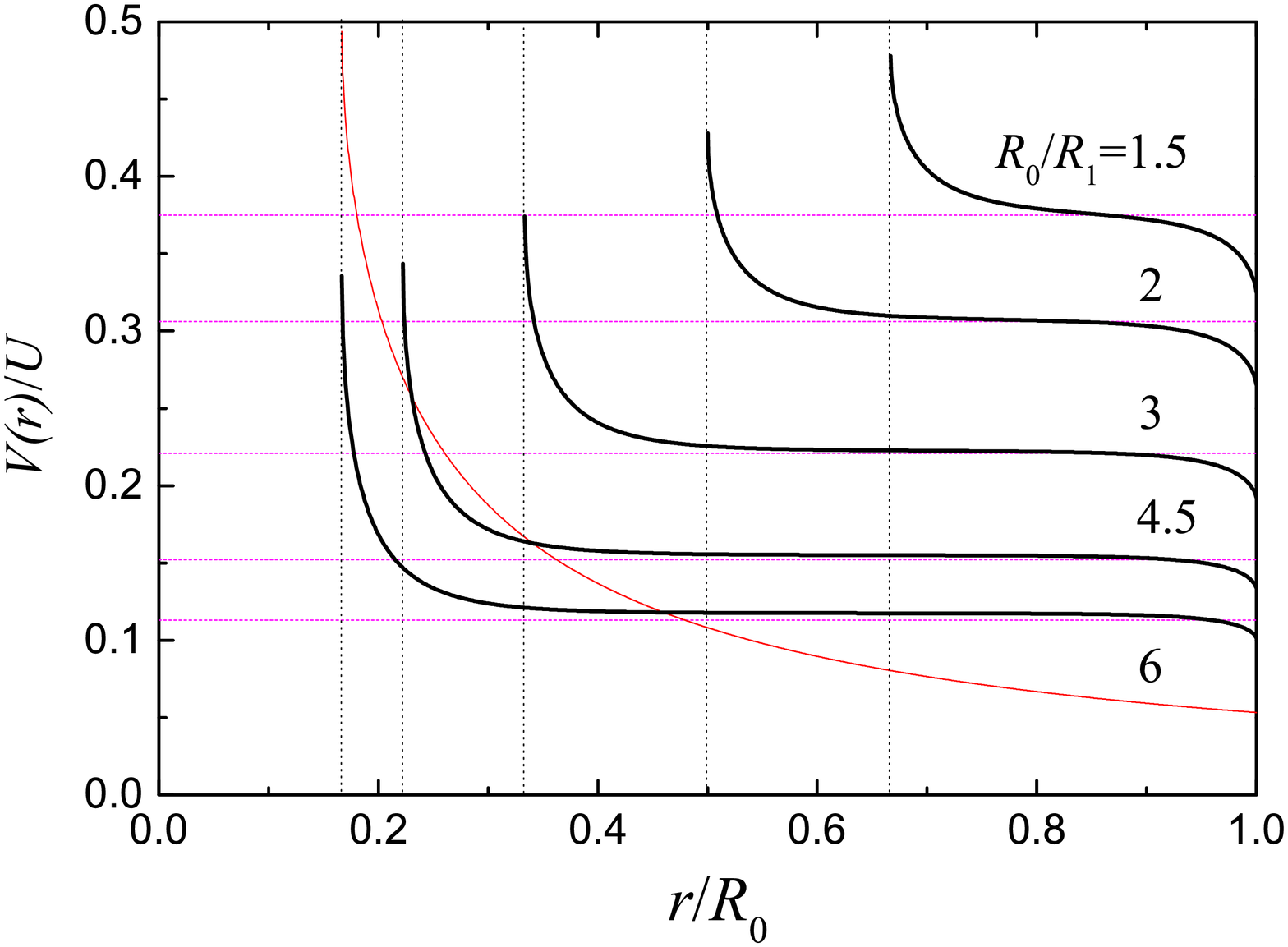}
\caption{\label{fig.5}(Color online) Bulk potential profiles for Corbino disk devices with 
different ratios $b=R_0/R_1$ from 1.5 to 6 for ${\rm K} \equiv l/R_0=0.1$ in the absence of 
momentum-relaxing scattering. For comparison, thin red line shows the potential profile for 
purely ballistic transport, $V(r)=(U/\pi)\arcsin(R_1/r)$, at $b=6$. The horizontal lines 
show the magnitude of the flat part $V_{bulk}$ according to Eq. (36). The vertical lines 
indicate the positions of the inner boundary, $r=R_1$.}
\end{figure}
 
The potential profiles for Corbino disks with different $b$, calculated at ${\rm K}=0.1$ in 
the absence of momentum-relaxing scattering, when ${\cal R}={\cal R}^*$, are shown in Fig. 5. 
They demonstrate almost flat electrochemical potentials in the middle regions and rapid 
change of the potentials in the Knudsen layers of width $\simeq l$ near the boundaries. 
The magnitudes of the flat parts of the potentials decrease with increasing $b$ and are in
good agreement with those given by Eq. (36). 
If $R_0-R_1 \ll R_0$, the behavior of the 
potential in Corbino devices approaches to that shown in Fig. 4, since the curvature 
effect becomes no longer important. To account for the momentum-relaxing scattering, one should 
add the contribution $I \ln(R_0/r)/2 \pi \sigma_0=I G_S^{-1} (2 R_1/\pi l_{tr}) \ln(R_0/r)$ 
to the potential profile. The results presented are consistent with the data obtained 
by experimental imaging of the potential distribution in the Corbino device \cite{kumar}.  

\begin{figure}[ht!]
\includegraphics[width=9.cm,clip=]{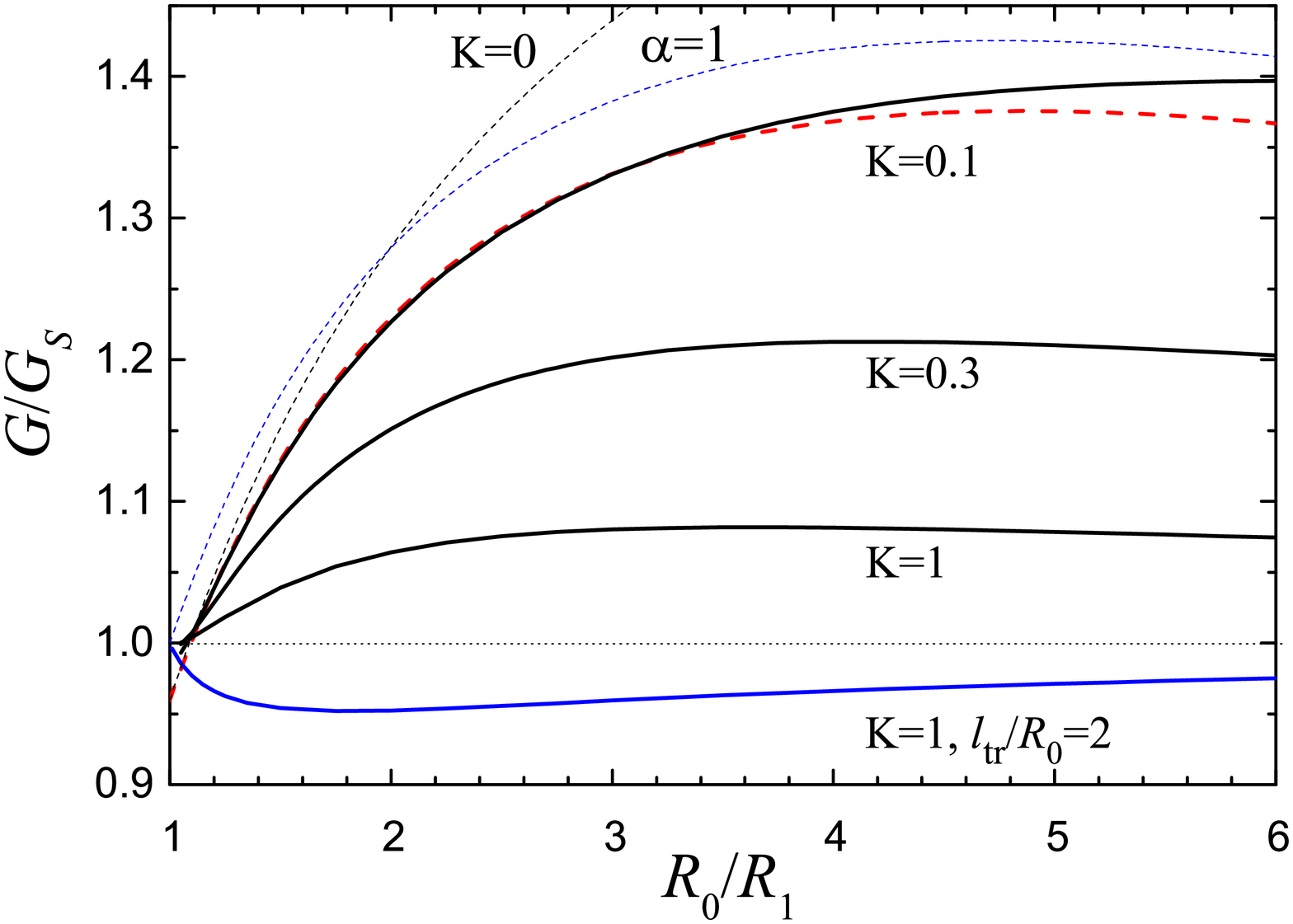}
\caption{\label{fig.6}(Color online) Conductance of Corbino disk device as a function of the 
ratio $R_0/R_1$. Three upper plots (black solid lines) are calculated in the absence of momentum-relaxing 
scattering, $l_{tr} \rightarrow \infty$, and the lower plot corresponds to $l_{tr}=l_e=2l=2 R_0$. 
The dashed lines show the approximate results according to Eq. (33): ${\rm K}=0.1$ and 
$\alpha=0.96$ (red); ${\rm K}=0.1$ and $\alpha=1$ (blue); ${\rm K}=0$ and $\alpha=0.96$ (black).} 
\end{figure}

The dependence of the conductance on the ratio $R_0/R_1$ is shown in Fig. 6. In the absence of 
momentum-relaxing scattering, when ${\cal R}={\cal R}^*$, the numerical result at ${\rm K}=0.1$ is in 
good agreement with the approximate result given by Eq. (33). The deviation in the region of large 
$R_0/R_1$ occurs because the curvature-induced correction is no longer small, $l/R_1= b {\rm K} \sim 1$, 
and Eq. (33) loses its validity. Thin dashed lines in Fig. 6 show the results of more crude 
approximations that neglect either the effect of scattering in the Knudsen layer ($\alpha=1$) 
or the curvature effect ($C=0$ in Eq. (28) or, equivalently, ${\rm K}=0$ in Eq. (33)). In 
general, the conductance is the highest when the momentum-conserving scattering is 
strong and the momentum-relaxing scattering is weak. The superballistic conductance $G > G_S$ does 
not require the hydrodynamic transport regime and can be observed even at $l_e > R_0$ if the
momentum-relaxing scattering is weak enough. To obtain a considerable superballistic effect, 
it is preferable to use the devices with $R_0/R_1>2$.

\begin{figure}[ht!]
\includegraphics[width=9.cm,clip=]{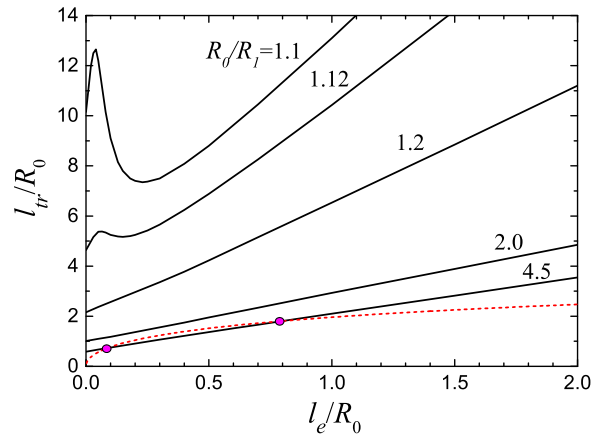}
\caption{\label{fig.7}(Color online) The borders of superballistic behavior for Corbino 
disks with different ratios $R_0/R_1$ (indicated) in the space of parameters $l_e$ and 
$l_{tr}$. The lines correspond to $G=G_S$. In the regions above the lines the transport 
is superballistic, $G>G_S$. Red dashed line shows the dependence of $l_{tr}/R_0$ on 
$l_e/R_0$ when temperature is changed, plotted by using the experimental details 
of Ref. \cite{kumar}. The intersection points of this plot with the line $G=G_S$ 
at $R_0/R_1=4.5$ correspond to the temperatures 31 K and 93 K.}
\end{figure}

The important question about the existence of superballistic conductance depending on the parameters 
of the problem is addressed in more detail in Fig. 7. It shows the lines separating the regimes $G>G_S$ 
and $G<G_S$ in the parameter space. These lines are nearly straight if $R_1$ is not too close to $R_0$. 
Otherwise, when $b-1$ is comparable to $1-\alpha$, the shape of the border lines is essentially 
different from linear at small $l_e/R_0$. Regardless of the value of $R_0/R_1$, the superballistic 
conductance requires $l_{tr}>l_e$, that is the rate of electron-electron scattering must be higher than 
the rate of momentum-relaxing one. In the hydrodynamic limit, $l_e/R_0 \rightarrow 0$, the superballistic 
conductance requires $l_{tr}/R_0 > (2/\pi) \ln b/[b-(b+1)/2 \alpha]$ and exists at $b>1/(2\alpha-1)$. 
For finite $l_e$, however, the superballistic conductance may exist even at $b<1/(2\alpha-1)$, although 
in this case the conductance only slightly exceeds $G_0$. The characteristic lengths $l_{tr}$ 
and $l_e$ are varied by the temperature $T$. For a degenerate fermion gas, 
\begin{eqnarray}
l_e \simeq \gamma v_F \hbar \varepsilon_F/T^2, 
\end{eqnarray}  
where $\gamma$ can be approximated by a numerical constant of order unity. On the other hand, 
$l_{tr} \sim 1/T$ if the main mechanism of momentum-changing scattering is the interaction of 
electrons with acoustic phonons and $T$ exceeds Bloch-Gr\"uneisen temperature. Therefore, by 
changing the temperature one moves in the parameter space along the square root line $l_{tr} \sim \sqrt{l_e}$, 
which may intersect the border lines shown in Fig. 7 in two points. This means that the 
superballistic conductance is expected to exist in a temperature interval corresponding to 
the interval between these points. In the presence of a considerable electron-electron 
scattering, the resistance should have a minimum as a function of $T$ that 
correlates with Eq. (39) and depends on the system size, like in the Gurzhi effect \cite{gurzhi}. Such 
a behavior was recently observed \cite{kumar} in graphene Corbino disks. Using the device dimensions 
$R_0=9$ $\mu$m and $R_1=2$ $\mu$m, density $n=4.5 \times 10^{11}$ cm$^{-2}$, graphene Fermi velocity 
$v_F=10^8$ cm/s, and the temperature dependence $l_{tr}$[$\mu$m]$=(0.016+0.0015 T [{\rm K}] )^{-1}$ 
extracted from the experimental data \cite{kumar}, one may plot the dependence of $l_{tr}$ on $l_{e}$. 
This dependence, obtained with $\gamma=1$ in Eq. (39), is also shown in Fig. 7. According to the 
calculations, the superballistic transport should be observed in the temperature interval between 
31 K and 93 K with the minimum of the resistance near 60 K, which is consistent with the 
experimental results of Ref. \cite{kumar}.

\section{Summary}

In this work, the classical transport properties of electron systems contacted to perfectly conducting 
electrodes (leads) have been studied. If the electron gas in the bulk is in the hydrodynamic transport 
regime, there exists a thin Knudsen layer separating the lead from the hydrodynamic electron state 
[Fig. 1 (b)]. The interface between the lead and this state can be characterized by the conductance 
per unit contact area, ${\cal G}$, which is intrinsically superballistic: ${\cal G}=2 \alpha {\cal G}_S$, 
where ${\cal G}_S$ is the Sharvin conductance per unit area and $\alpha$ is a numerical constant 
slightly smaller than unity. Remarkably, within the elastic relaxation-time approximation for 
electron-electron collision integral, which is often applied in theoretical description of electron 
transport, $\alpha$ depends only on the dimensionality of the system and is approximately equal to 
0.96 for 2D systems and 0.94 for 3D systems. Application of the modified double-time approximation 
for electron-electron collision integral \cite{gurzhi1}, which takes into account the difference in 
relaxation times for even and odd angular harmonics of the distribution function, brings $\alpha$ 
closer to unity (Fig. 3). The correction to ${\cal G}$ due to finite curvature of the contact boundary 
also has been found. It is shown that ${\cal G}$ decreases for the convex boundary and increases 
for the concave boundary. 

Based on these results, the conductance of highly symmetric 2D microstructures [Fig. 1 (c,d)], 
where the current is always normal to the contact boundaries, has been described. The important example 
of this kind is the Corbino disk device, whose conductance $G$ can exceed the Sharvin conductance 
$G_S$ owing to the superballistic contact property of the device boundaries. The conductance of 
Corbino disks has been calculated as well by means of numerical solution of the Boltzmann kinetic 
equation, which describes different transport regimes and transitions between them, depending on 
the characteristic lengths of the problem. In the hydrodynamic transport regime, when 
the Knudsen number ${\rm K}$ is small, the calculated conductance and the electrochemical 
potential in the bulk are in good agreement with the simple analytical formulas obtained from 
the consideration of a single contact boundary (Figs. 5 and 6). The kinetic equation approach 
shows that the superballistic conductance of Corbino disks does not require the hydrodynamic 
transport regime, although the domination of electron-electron scattering over the momentum-relaxing 
scattering is a necessary requirement. The map in the parameter space, indicating the regions 
with superballistic conductance $G > G_S$, has been presented (Fig. 7). By considering the 
dependence of electron-electron and momentum-relaxing mean free path lengths on temperature, 
it is concluded that the superballistic conductance in Corbino disks can be observed within a 
certain interval of temperatures, described in agreement with experimental results \cite{kumar}. 

The calculation of the conductance standing in Eq. (1) is important by itself, without regard 
to the specific problem of superballistic transport. Indeed, Eq. (1) relates the normal current 
density to the voltage of the contact and should be considered as a hydrodynamic boundary 
condition for the normal current density at the current-penetrable boundary. On the other 
hand, the tangential current density ${\bf j}_t$ is related to its normal derivative 
at the boundary by Maxwell's boundary condition, ${\bf j}_t=l_S \nabla_n {\bf j}_t$ 
\cite{jensen},\cite{kiselev},\cite{raichevu}, which is usually applied to the hard-wall 
boundaries, where the normal current is zero. 
The tangential current density at the current-penetrable boundaries can be as well described 
by Maxwell's boundary condition derived in the fully diffuse limit, which corresponds to the 
slip lengths $l_S=0.582 l_e$ and $l_S=0.637 l_e$ for degenerate 3D and 2D fermion gases, 
respectively \cite{raichevu}. By combining Maxwell's boundary conditions with Eq. (1), 
one obtains a full set of hydrodynamic boundary conditions, describing both tangential and 
normal current for both hard-wall and current-penetrable boundaries. These boundary conditions, 
together with the Navier-Stokes equation, form a Cauchy problem for determination of the current 
density and potential distribution in various microstructures in the hydrodynamic transport regime, 
which is important for applications, in particular, for development of viscous electronics.

\begin{appendix}

\section{}

The solution of Eq. (14) with $l_e=l$ and boundary conditions Eq. (30) is
\begin{eqnarray}
g^+_{\varphi}(y)=g_y (1-e^{-y/l \sin \varphi} ) \sin \varphi 
+ \frac{eU}{2} e^{-y/l \sin \varphi} \nonumber \\
+ \int_{0}^{y} \frac{d y'}{l \sin \varphi} e^{(y'-y)/l \sin \varphi} {\overline g}(y'),
\end{eqnarray}
\begin{eqnarray}
g^-_{\varphi}(y)=-g_y(1-e^{(y-L)/l \sin \varphi} ) \sin \varphi 
- \frac{eU}{2}e^{(y-L)/l \sin \varphi} \nonumber \\
+ \int_{y}^{L} \frac{d y'}{l \sin \varphi} e^{(y-y')/l \sin \varphi} {\overline g}(y'). ~~~~~~
\end{eqnarray}
With ${\overline g}(y)=\overline{[( g^+_{\varphi}+g^-_{\varphi} )]}_+=(2\pi)^{-1}\int_0^{\pi} d \varphi 
[ g^+_{\varphi}(y)+g^-_{\varphi}(y) ]$ and 
$g_y = 2 \overline{[( g^+_{\varphi} -g^-_{\varphi} ) \sin \varphi ]}_+=\pi^{-1}\int_0^{\pi} d \varphi 
[ g^+_{\varphi}(y)-g^-_{\varphi}(y) ] \sin \varphi$, one gets an integral 
equation for ${\overline g}(y)$. Since ${\overline g}(y)=eV(y)$, this equation is 
written as an equation for the electrochemical potential: 
\begin{eqnarray}
V({\tilde y}) = \frac{U}{2} {\cal S}({\tilde y}) + \int_0^{L/l} d {\tilde y} {\cal K}({\tilde y},{\tilde y}') 
V({\tilde y}'),
\end{eqnarray}
where the dimensionless coordinates ${\tilde y} = y/l$ and ${\tilde y}' =y'/l$ are used. In Eq. (A3),
\begin{eqnarray}
{\cal S}({\tilde y}) = \zeta^-_0({\tilde y})- \zeta^-_1({\tilde y})\zeta^+_1({\tilde y})/\zeta^+_2({\tilde y}), 
\end{eqnarray}
\begin{eqnarray}
{\cal K}({\tilde y},{\tilde y}')= K_0({\tilde y},{\tilde y}') -  
\frac{\zeta^-_1({\tilde y})}{\zeta^+_2({\tilde y})} K_1({\tilde y},{\tilde y}'),
\end{eqnarray}
$K_0$ and $K_1$ are given by Eq. (18), and 
\begin{eqnarray}
\zeta^{\pm}_k({\tilde y})= \overline{[\sin^k \varphi (e^{-{\tilde y}/\sin \varphi} \pm e^{({\tilde y}-L/l)/
\sin \varphi})]}_+ .
\end{eqnarray}
Once $V(y)$ is found, the ratio $G^*/G_S$ is found from the expression
\begin{eqnarray}
\frac{G^*}{G_S}= \frac{\pi \zeta^+_1(0)}{4 \zeta^+_2(0)} \left[1 - \frac{2}{\zeta^+_1(0)} \int_0^{L/l} d \tilde{y} 
\zeta_0({\tilde y}) V({\tilde y})/U \right].
\end{eqnarray}
where $\zeta_0$ is given by Eq. (19). In the ballistic limit, $l/L \rightarrow \infty$, $G^*=G_S$. 
In the hydrodynamic limit, $l/L \rightarrow 0$, one has $\zeta^{\pm}_k=\zeta_k$, ${\cal S}=S$, 
${\cal K}=K$, and Eq. (A3) becomes identical to Eq. (15) with $\Delta U=U/2$, reflecting the 
property that half of the total potential drops near each boundary. Since $\zeta_1(0)=
1/\pi$ and $\zeta_2(0)=1/4$, Eq. (A7) is reduced to $G^*/G_S=\alpha$, where $\alpha$ is 
given by Eq. (20).

\section{}

The solution of Eq. (37) with $l_e=l$ and boundary conditions Eq. (38) is 
\begin{eqnarray}
g^+_{\varphi}(\rho) = eU \theta(\rho_1-|w|) e^{\psi_{w}(\rho_1)-\psi_{w}(\rho)} +\theta(|w|-\rho_1) \nonumber \\
\times \int_{{\rm max}\{\rho_1,|w|\}}^{\rho_0} d \rho' e^{-\psi_{w}(\rho)-\psi_{w}(\rho')} 
\left[\frac{\rho'{\overline g}(\rho')}{\psi_w(\rho')} -\frac{A}{l\rho'}\right] \nonumber \\
+\int_{{\rm max}\{\rho_1,|w|\}}^{\rho} d \rho' e^{-\psi_{w}(\rho)+\psi_{w}(\rho')} 
\left[\frac{\rho' {\overline g}(\rho') }{\psi_w(\rho')} +\frac{A}{l\rho'}\right],
\end{eqnarray}
\begin{eqnarray}
g^-_{\varphi}(\rho)=\int_{\rho}^{\rho_0} d \rho' e^{\psi_{w}(\rho)-\psi_{w}(\rho')} 
\left[\frac{\rho' {\overline g}(\rho')}{\psi_w(\rho')} - \frac{A}{l\rho'}\right],
\end{eqnarray}
where the dimensionless quantities are introduced according to $\rho=r/l$, $\rho'=r'/l$, 
$\rho_0=R_0/l$, $\rho_1=R_1/l$, $w=\rho \cos \varphi$, and $\psi_{w}(\rho)
= \sqrt{\rho^2-w^2}$. The latter is related to $\varphi$ as $\psi_{w}(\rho)=
\rho \sin \varphi$. 
The solution satisfies the necessary requirements $g^+_{0}(\rho)=g^-_{0}(\rho)$ and 
$g^+_{\pi}(\rho)=g^-_{\pi}(\rho)$ expressing periodicity of $g_{\varphi}(\rho)$ and 
its continuity at $\varphi=\pi$. Applying ${\overline g}(\rho)=eV(\rho)=(2\pi)^{-1}\int_0^{\pi} 
d \varphi [ g^+_{\varphi}(\rho)+g^-_{\varphi}(\rho) ]$ and $A/r =\pi^{-1}\int_0^{\pi} d \varphi 
[ g^+_{\varphi}(\rho)-g^-_{\varphi}(\rho) ] \sin \varphi$, one obtains the integral equation
\begin{eqnarray}
V(\rho) = U {\cal L}(\rho) + \int_{\rho_1}^{\rho_0} d \rho' {\cal Q}(\rho,\rho') V(\rho').
\end{eqnarray}
The functions entering Eq. (B3) are
\begin{eqnarray}
{\cal L}(\rho)={\cal L}_0(\rho)+ {\cal L}_1(\rho) Z_0(\rho)/[1-Z_1(\rho)], \\
{\cal Q}(\rho,\rho')={\cal Q}_0(\rho,\rho')+ {\cal Q}_1(\rho,\rho')Z_0(\rho)/[1-Z_1(\rho)],
\end{eqnarray}
\begin{eqnarray}
{\cal L}_0(\rho)=\frac{1}{\pi} \int_0^{\rho_1} d w \frac{e^{\psi_{w}(\rho_1)-\psi_{w}(\rho)}}{\psi_{w}(\rho)}, \\ 
{\cal L}_1(\rho)=\frac{2}{\pi} \int_0^{\rho_1} d w e^{\psi_{w}(\rho_1)-\psi_{w}(\rho)},
\end{eqnarray}
\begin{eqnarray}
{\cal Q}_0(\rho,\rho')=\frac{1}{\pi}\int_0^{\rho_m} 
d w \frac{e^{-|\psi_{w}(\rho)-\psi_{w}(\rho')|} \rho'}{\psi_{w}(\rho) \psi_{w}(\rho')} \nonumber \\
+ \frac{1}{\pi} \int_{\rho_1}^{\rho_m} d w \frac{e^{-\psi_{w}(\rho)-\psi_{w}(\rho')}\rho'}{\psi_{w}(\rho) 
\psi_{w}(\rho')},
\end{eqnarray}
\begin{eqnarray}
{\cal Q}_1(\rho,\rho')=\frac{2}{\pi} \int_0^{\rho_m} 
d w \frac{e^{-|\psi_{w}(\rho)-\psi_{w}(\rho')|}{\rm sgn}(\rho-\rho') \rho'}{\psi_{w}(\rho')} \nonumber \\
+ \frac{2}{\pi} \int_{\rho_1}^{\rho_m} d w \frac{e^{-\psi_{w}(\rho)-\psi_{w}(\rho')}\rho'}{\psi_{w}(\rho')},~~~
\end{eqnarray}
\begin{eqnarray}
Z_0(\rho)=\frac{1}{\pi} \int_{\rho_1}^{\rho_0} \frac{d \rho'}{\rho'} \left[ \int_0^{\rho_m} d w \frac{e^{-|\psi_{w}(\rho)-\psi_{w}(\rho')|} }{\psi_{w}(\rho)}  
\right. \nonumber \\
\left. \times {\rm sgn}(\rho-\rho') - \int_{\rho_1}^{\rho_m} d w \frac{e^{-\psi_{w}(\rho)-\psi_{w}(\rho')}}{\psi_{w}(\rho)} \right],
\end{eqnarray}
and
\begin{eqnarray}
Z_1(\rho)=\frac{2}{\pi} \int_{\rho_1}^{\rho_0} \frac{d \rho'}{\rho'} \left[ \int_0^{\rho_m} d w e^{-|\psi_{w}(\rho)-\psi_{w}(\rho')|}  
\right. \nonumber \\
\left. - \int_{\rho_1}^{\rho_m} d w e^{-\psi_{w}(\rho)-\psi_{w}(\rho')} \right],
\end{eqnarray}
where $\rho_m={\rm min}\{\rho,\rho'\}$. The conductance $G^*=1/R^*$ is determined by the solution 
of Eq. (B3) as follows:
\begin{eqnarray}
\frac{G^*}{G_S}=\frac{1-(\pi/\rho_1) \int_{\rho_1}^{\rho_0} d \rho \rho
{\cal L}_0(\rho) V(\rho)/U} {1-\int_{\rho_1}^{\rho_0} d \rho {\cal L}_1(\rho)/\rho}.
\end{eqnarray}
The conductance depends on the ratios $R_0/l$ and $R_1/l$. In the ballistic limit, 
when $R_0/l \rightarrow 0$, the integral terms in Eq. (B12) go to zero as well and 
$G^*=G_S$, while Eq. (B3) is reduced to $V(\rho) = U {\cal L}_0(\rho)$ describing the 
potential distribution $V(r)=(U/\pi)\arcsin(R_1/r)$ \cite{shavit,kumar}. In the 
hydrodynamic limit, when $l/R_1 \rightarrow 0$, $l/(R_0-R_1) \rightarrow 0$, the 
relations $G^*/G_S = 2 b \alpha/(b+1)$ and $V(r)=V_{bulk}=U/(b+1)$ are restored. 

\end{appendix}

\end{document}